\documentclass[a4paper,11pt]{article}
\usepackage{pos}
\usepackage{here}
\usepackage{physics}
\usepackage{bm}
\usepackage[compat=1.1.0]{tikz-feynhand}
\usepackage[subrefformat=parens]{subcaption}
\usepackage{bxtexlogo}

\title{Parameter Optimization of Domain-Wall Fermion using Machine Learning}

\author*[a,b]{Shunsuke Yasunaga}
\author[a,b]{Kenta Yoshimura}
\author[c,b,d]{Akio Tomiya}
\author[e,f]{Yuki Nagai}

\affiliation[a]{Department of physics, Institute of Science Tokyo, Meguro, Tokyo 152-8550, Japan}
\affiliation[b]{RIKEN Center for Computational Science, Kobe, Hyogo 650-0047, Japan}
\affiliation[c]{Department of Information and Mathematical Sciences, Tokyo Woman’s Christian University, Suginami, Tokyo 167-8585, Japan}
\affiliation[d]{Department of Physics, Kyoto University, Kyoto, Kyoto 606-8502, Japan}
\affiliation[e]{Information Technology Center, The University of Tokyo, Kashiwa, Chiba 277–0882, Japan}
\affiliation[f]{Department of Advanced Materials Science, The University of Tokyo, Kashiwa, Chiba 277–8561, Japan}

\abstract{We study a parameter optimization of domain-wall fermions to improve chiral symmetry based on machine learning. Domain-wall fermions involve coefficients along the fifth dimension, which can be treated as trainable parameters to reduce the chiral symmetry violation caused by the finite extent of the fifth dimension. As the loss function, we use the residual mass estimated stochastically on a single gauge configuration. Numerical tests on a $L^3\times T\times L_5=4^3\times8\times8$ lattice demonstrate the feasibility of this framework.}

\FullConference{
  The 42nd International Symposium on Lattice Field Theory (LATTICE2025)\\
  3-8 November 2025\\
  Mumbai, India
}

\begin{document}
\maketitle
\section{Introduction}
Chiral symmetry and its spontaneous breaking are fundamental concepts in quantum chromodynamics (QCD). However, in the lattice formulation, which provide one of the most promising regularizations of nonperturbative field theory, cannot preserve chiral symmetry in a straightforward way because of the Nielsen-Ninomiya theorem~\cite{nielsen}. This obstruction motivates some descretizations of the lattice fermions. One powerful approach is the domain-wall fermion formulation, which realizes good chiral symmetry on the lattice by introducing an additional fifth dimension. In the limit where the extent of the fifth dimension becomes infinite, domain-wall fermions satisfy the Ginsparg-Wilson relation~\cite{ginsparg} and possess an exact lattice chiral symmetry even at finite lattice spacing.

Domain-wall fermions were originally proposed as a formulation of chiral fermions on the lattice, motivated by chiral gauge theories as in the Standard Model~\cite{kaplan}. Subsequently, their application for lattice QCD with chiral symmetry was developed~~\cite{shamir,furman}, and nowadays domain-wall fermions are widely employed as a chiral symmetric discretization in the lattice QCD. Tuning of domain-wall coefficients has been explored as a practical way to improve chiral symmetry in domain-wall fermions. Dirac operators satisfying the Ginsparg-Wilson relation can be expressed in terms of the sign function~\cite{neuberger}. The Zolotarev approximation provides a rational approximation to the sign function and enables a highly accurate realization of chiral symmetry by optimization of the coefficients appearing in the numerator~\cite{chiu,chen}. In contrast, the M\"obius domain-wall formulation introduces a transformation in the denominator as well~\cite{brower}. Furthermore, an extension to complex coefficients, known as zM\"obius, has also been developed~\cite{abramczyk}.

Machine learning has recently been explored in lattice QCD as a general framework to improve the efficiency of numerical simulations~\cite{tomiya,lawrence}. Because it provides a flexible way to handle complicated optimization problems in high-dimensional parameter spaces, it may also be useful for tuning internal parameters of lattice actions. Nevertheless, applications aimed at improving physical properties have so far remained relatively limited. For domain-wall fermions, the coefficients in the fifth-dimensional structure control the approximation to the sign function and hence the quality of chiral symmetry. This feature makes them a natural target for optimization. In this work, we investigate their optimization using a machine-learning-based framework.


\section{Method}
\subsection{Parameters of domain-wall fermion}
The five-dimensional domain-wall Dirac operator $D_5$ is written as
\begin{equation}
    (D_5)_{nm,\,st} = (D_\text{W})_{nm}(b_s\delta_{st}+c_sF_{st})+\delta_{nm}(\delta_{st}-F_{st}),\label{equ:d5}
\end{equation}
where the subscripts $n,m$ denote four-dimensional lattice sites, and $s,t=1,\ldots,L_5$ label the slices in the fifth dimension. The operator $D_\text{W}$ denotes the Wilson--Dirac operator, while the matrix $F_{st}$ represents hopping in the fifth dimension and implements the fermion with chiral symmetry on the boundaries of the fifth dimension. The coefficients $b_s$ and $c_s$ control the mapping from the five-dimensional formulation to the corresponding effective four-dimensional operator. In practice, the effective four-dimensional operator can be expressed in terms of a product of transfer matrices along the fifth dimension. This product encodes the approximation to the sign function, and thus determines the quality of chiral symmetry. The transfer matrix is constructed from the Hermitian operator $H_s$ defined by
\begin{equation}
    H_s = \gamma_5\frac{(b_s+c_s)D_\text{W}}{2+(b_s-c_s)D_\text{W}}.\label{equ:kernel}
\end{equation}
Typically, the coefficients $b_s$ and $c_s$ are taken to be independent of $s$; in this case they are referred to as the M\"obius parameters~\cite{brower}. One may also allow only the combination $b_s+c_s$ in the numerator of Eq.~\eqref{equ:kernel} to depend on $s$. This makes it possible to improve the chiral symmetry by optimizing the rational approximation to the sign function, as in the Zolotarev approximation~\cite{chiu,chen}. In this work, we generalize the coefficients by allowing both $b_s$ and $c_s$ to depend on $s$, and treat them as independent parameters on each slice in the fifth dimension.

\subsection{Residual mass}
The violation of chiral symmetry due to lattice artifacts is quantified by the residual mass $m_\text{res}$. From the axial Ward--Takahashi identity, $m_\text{res}$ can be defined as the plateau value of the ratio of appropriate pseudoscalar correlation functions~\cite{aoki,cppacs,blum}.
It is also useful to introduce the "global" residual mass, which can be evaluated without identifying a plateau in Euclidean time. Following Ref.~\cite{chen}, the global residual mass can be expressed in terms of the effective four-dimensional operator $D_4$ as
\begin{equation}
    m_\text{res} = \frac{\text{Re}\expval{\Tr D_4^{-1}}}{\expval{\Tr (D_4^{\dag}D_4)^{-1}}} - m_q\label{equ:global},
\end{equation}
where $m_q$ is the input bare quark mass. The traces are taken over the four-dimensional lattice sites as well as spin and color indices. Note that these two definitions can differ in the presence of near-zero modes~\cite{cossu}. The effective four-dimensional operator $D_4$, with the contact term subtracted, can be obtained from the boundary of the fifth dimension as~\cite{edwards}
\begin{equation}
    D_4^{-1} = \left[P^\dag D_5^{-1}RP\right]_{(s,t)=(1,1)},
\end{equation}
where $P$ ($P_{st}=P_L\delta_{st}+P_R\delta_{s+1,t}$ for $s\neq L_5$ and $P_{L_5,t}=P_L\delta_{t,L_5}+P_R\delta_{t,1}$) and $R$ ($R_{st}=\delta_{s,L_5+1-t}$) denote the chiral projection and reflection operators, respectively.

\subsection{Loss function and its deriavtive}
Machine learning provides a systematic optimization framework in which one specifies a loss (objective) function and iteratively updates the model parameters to minimize it. In gradient-based methods, the parameters are updated using the gradient of the loss with respect to those parameters.

To apply this framework to domain-wall fermions, we treat the domain-wall coefficients $b_s$ and $c_s$ as parameters to be optimized and choose a loss function that quantifies the violation of chiral symmetry. As a per-configuration loss, we use the global residual mass in Eq.~\eqref{equ:global} evaluated on a single gauge configuration:
\begin{equation}
    \mathcal{L} = \mathcal{M}^2 = \left(\frac{\text{Re}\Tr D_4^{-1}}{\Tr (D_4^{\dag}D_4)^{-1}}-m_q\right)^2.
\end{equation}
Although one can take a configuration average using a mini-batch of gauge configurations, in this work we update the parameters on a per-configuration basis.

Next, we calculate the gradient of the loss function with respect to the fifth-dimensional vectors $b_s$ and $c_s$. Using a set of noise vectors $\eta_k$, the numerator and denominator of the loss function are rewritten as stochastic estimators
\begin{equation}
    \mathcal{M} = \mathcal{\frac{N}{D}}
    =\frac{\text{Re}\left(\sum_k\eta_k^\dag P^\dag D_5^{-1}RP \eta_k\right)}{\sum_k \eta^\dag_k \left(D_5^{-1}RP\right)^\dag Q \left(D_5^{-1}RP\right) \eta_k},
\end{equation}
where $Q=\text{diag}(P_L,0,\ldots,0,P_R)$ is a projection operator which acts only on the boundaries of fifth dimension. The derivatives with respect to the domain-wall parameters follow from Eq.~\eqref{equ:d5}:
\begin{align}
    \pdv{(D_5)_{tu}}{b_s} = \delta_{st}\delta_{tu}D_\text{W},\\
    \pdv{(D_5)_{tu}}{c_s} = \delta_{st}F_{tu}D_\text{W}.
\end{align}
After summing over the index $t$, these derivatives can be regarded as the $(s,u)$ components of a matrix acting in the fifth-dimensional space. Since the $t$ index is contracted, the matrix products surrounding $\grad_{b,c}D_5$ can be separated into left and right factors. Then, the gradients of $N$ and $D$ can be written as
\begin{align}
    \grad_{b,c}\mathcal{N} &= \text{Re}\sum_k\left[\left(D_5^\dag\right)^{-1}P\eta_k\right]^*\odot\left[\left(\grad_{b,c} D_5\right) \chi_k\right],\\
    \grad_{b,c}\mathcal{D} &= 2\text{Re}\sum_k\left[\left(D_5^\dag\right)^{-1}Q\chi_k\right]^*\odot\left[\left(\grad_{b,c} D_5\right) \chi_k\right],
\end{align}
where $\odot$ denotes the Hadamard product of vectors in the fifth dimension, and $\chi_k=D_5^{-1}RP \eta_k$.

\section{Numerical setup}

We compute parameter updates using the adaptive moment estimation (Adam) optimizer implemented in \texttt{Optimisers.jl}, with the learning rate set to $10^{-2}$.
Both the loss function and its derivatives are evaluated using ten independent noise vectors for each learning epoch.
The four-dimensional lattice size is $L^3\times T=4^3\times 8$, and we use the Wilson gauge action at $\beta=6.0$.
For the domain-wall fermion, we use a bare quark mass $m_q a=0.02$ and Pauli--Villars mass $m_\text{PV}=1.0$, domain-wall height $M_5=1.9$ ($\kappa=0.2381$), and fifth-dimensional extent $L_5=8$.
Simulations and measurements are carried out using \texttt{LatticeQCD.jl}, which we extended to implement residual-mass measurement, gradient computation, and parameter updates.
In the present setup, parameter updates and measurements are performed alternately for each gauge configuration; namely, after generating a configuration, we evaluate the loss and its gradient on that configuration and then update the parameters before proceeding to the next configuration.

We consider two parameterizations of the domain-wall kernel along the fifth direction, namely the \textit{M\"{o}bius} setting and the \textit{general} setting.
In the M\"{o}bius setting, the coefficients are taken to be uniform in the fifth dimension.
On the other hand, in the general setting, independent parameters are assigned to each fifth-dimensional slice, so that the parameter space is a $2L_5$-dimensional space spanned by $\{b_s, c_s\}$.
Unless otherwise stated, the initial parameters are set to $b_s = 1.5$, $c_s = 0.5$, which corresponds to the so-called scaled-Shamir kernel.

\section{Preliminary results}
\subsection{Decreasing of the loss function}

\begin{figure}[h]
    \centering
    \includegraphics[width=0.7\linewidth]{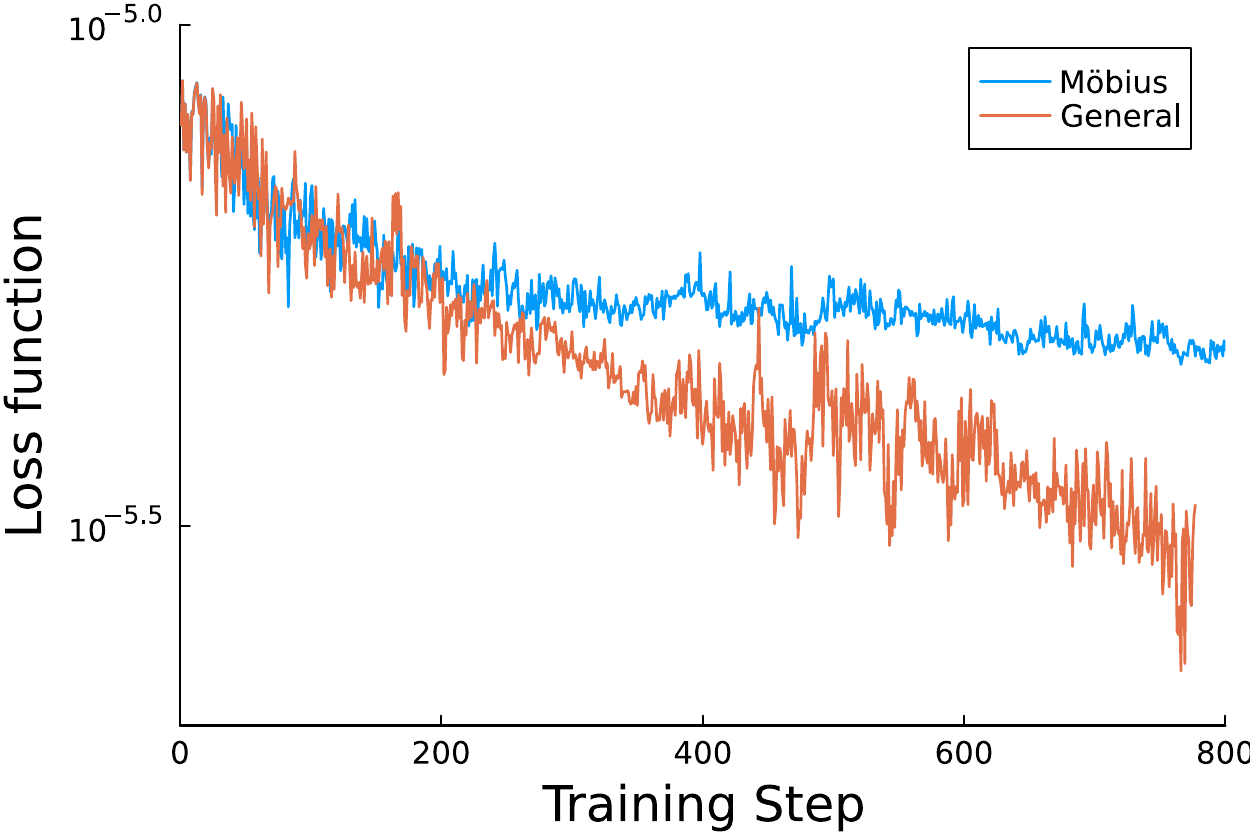}
    \caption{Training history of the loss function for the M\"{o}bius and general settings.}
    \label{fig:comp_loss}
\end{figure}

Figure~\ref{fig:comp_loss} shows the evolution of the loss function as a function of the training step for the M\"{o}bius and general settings.
In both cases, the loss decreases stably as the optimization proceeds, indicating that the training behaves as intended and improves the target quantity associated with chiral symmetry.
A comparison between the general and M\"{o}bius settings reveals that allowing independent parameters for each fifth-dimensional slice leads to a systematically smaller residual mass $M_{\mathrm{res}}$ than imposing a uniform value across all slices.
In other words, within the $2L_5$-dimensional parameter space spanned by $\{b_s, c_s\}$, there exist configurations that achieve better chiral symmetry than those lying on the M\"{o}bius constraint surface.

\begin{figure}[tp]
    \centering
    \includegraphics[width=0.5\linewidth]{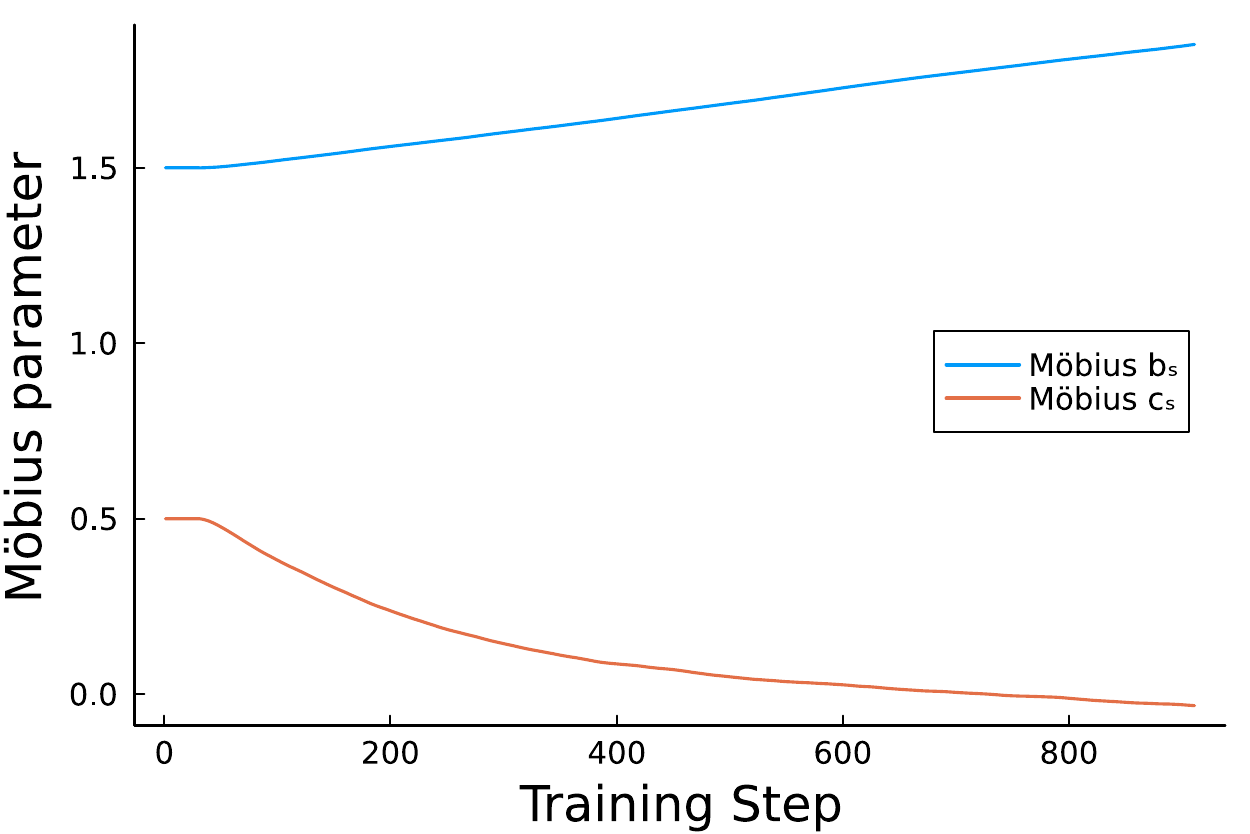}
    \caption{Training history of the M\"{o}bius parameters $b$ and $c$.}
    \label{fig:mobius_bc}
\end{figure}

\begin{figure}[tp]
    \centering
    \begin{minipage}[b]{0.49\textwidth}
        \includegraphics[width=\textwidth]{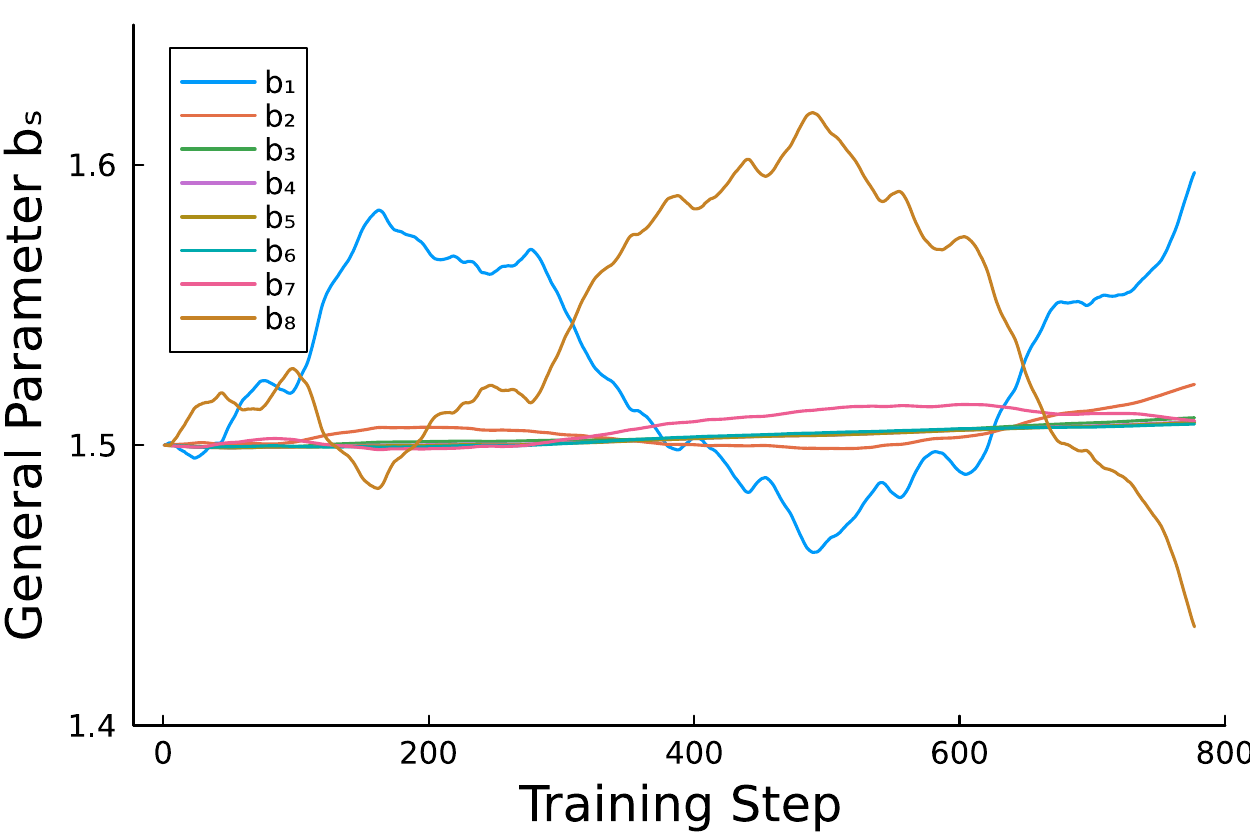}
    \end{minipage}
    \hfill
    \begin{minipage}[b]{0.49\textwidth}
        \includegraphics[width=\textwidth]{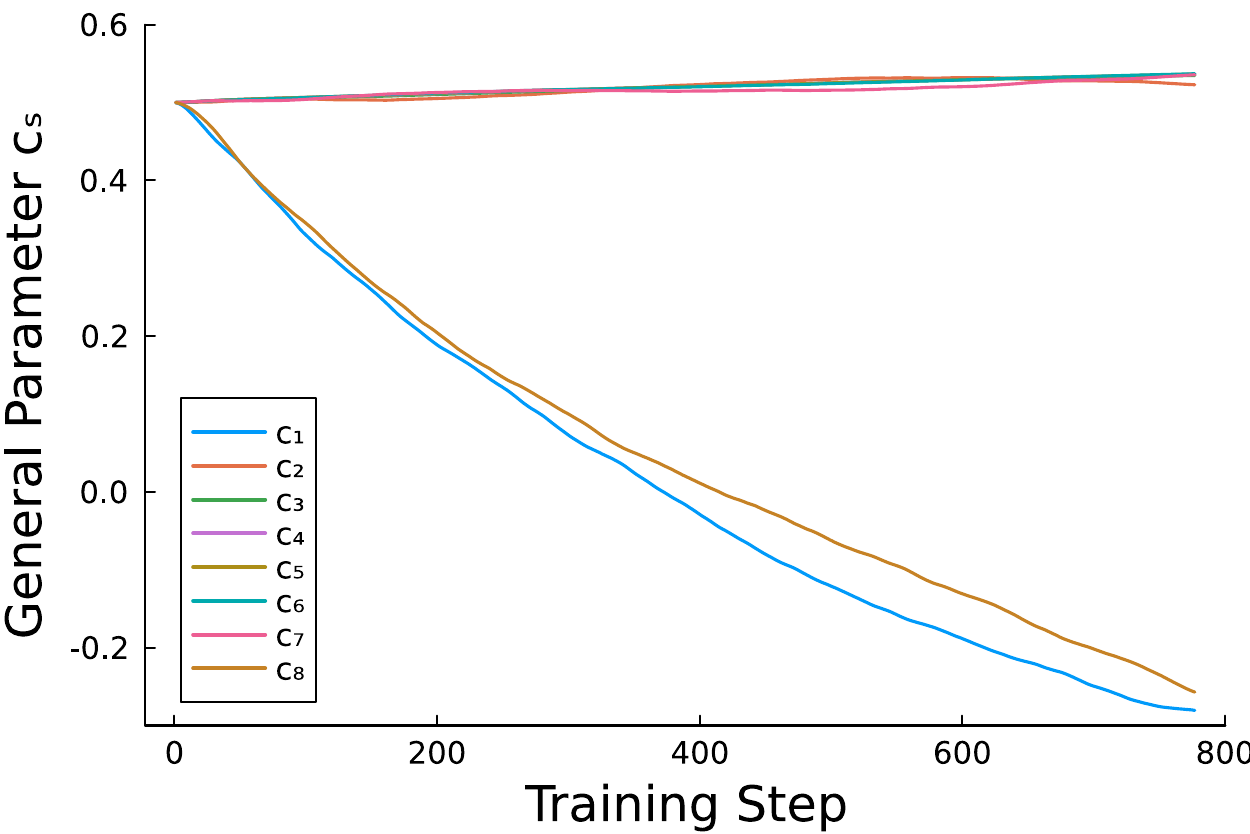}
    \end{minipage}
    \caption{Training histories of the slice-dependent domain-wall parameters $b_s$ (left) and $c_s$ (right) in the general setting.}
    \label{fig:general_bscs}
\end{figure}

\subsection{Evolution of the optimized parameters}

Figures~\ref{fig:mobius_bc} and \ref{fig:general_bscs} summarize the evolution of the optimized parameters in the M\"{o}bius and general settings.
In Fig.~\ref{fig:mobius_bc}, the two M\"{o}bius parameters $b$ and $c$ are displayed in a single panel.
In contrast, Fig.~\ref{fig:general_bscs} shows, in the left panel, each fifth-dimensional component of $b_s$, while the right panel is dedicated to $c_s$ in the general setting.
From Fig.~\ref{fig:general_bscs}, two characteristic features are observed.
Firstly, the dominant variations occur near the boundaries of the fifth dimension, i.e., at $s=1$ and $s=L_5$, whereas the bulk parameters remain comparatively stable throughout the optimization.
This behavior is consistent with patterns known from Zolotarev-type optimizations, where endpoint coefficients play a central role, and it suggests that the boundary slices control most of the effective freedom relevant for improving chiral symmetry.
Secondly, $b_s$ and $c_s$ exhibit qualitatively different convergence behavior.
In contrast to endpoint components of $b_s$ fluctuating during training, $c_s$ shows an approximately monotonic drift over the explored training range and does not exhibit clear saturation, even when the training is extended.
This contrast suggests that the loss function is more sensitive to variations in $b_s$ than in $c_s$.
In particular, the parameter subspace associated with $c_s$ may contain relatively flat directions, along which the loss changes only weakly, leading to a slow drift rather than rapid convergence.
Additionally, we find that highly negative $c_s$ values lead to convergence instabilities of the conjugate-gradient (CG) solver for the Dirac operator.
This indicates that, for a stable learning scheme, it is necessary to amend the loss function by adding constraint terms that moderate the variation of the $c_s$ parameters.

\section{Summary and outlook}
We studied parameter optimization of domain-wall fermions to improve chiral symmetry using machine learning. By adopting a stochastic estimator of the global residual mass as the loss function, we demonstrated that the loss can be reduced through parameter updates.

In future work, we will validate whether the obtained parameters effectively reduce the residual mass by measuring the conventional plateau definition on larger lattice volumes. A systematic study of the lattice-volume dependence, as well as an investigation of the behavior toward the continuum limit by varying the lattice spacing, would also be important.

We have also observed that changing the parameters modifies the convergence behavior of the solver, suggesting that these parameters may be tuned as a precondition to improve the solver conditioning. In such a scenario, multiple evaluations such as chiral-symmetry violation and solver performance are required, and we expect the machine-learning framework to work effectively for this multiobjective optimization.

\acknowledgments
We would like to thank Y. Aoki, I. Kanamori, K. Nakayama, and K. Nitadori for their continued discussions.
This work was supported by MEXT as Program for Promoting Researches on the Supercomputer Fugaku (Simulation for basic science: approaching the new quantum era; Grant Number JPMXP1020230411) and used computational resources of the Fugaku supercomputer provided by the RIKEN Center for Computational Science Project Numbers hp240295 and hp250224. We also used the Yukawa-21 supercomputer at the Yukawa Institute for Theoretical Physics, Kyoto University. S.Y. was supported by JST SPRING Grant Number JPMJSP2180. K.Y. was supported by JSPS KAKENHI Grant Number JP24KJ1110. A. T. was supported by JSPS KAKENHI Grant Numbers JP22H05111, JP22H05112, JP22K03539, JP25K07287, and by JST BOOST Grant Number JPMJBY24F1. Y. N. was supported by JSPS KAKENHI Grant Numbers JP22K03539, JP22K12052, JP22H05111, JP22H05114, and JP25K07287.

\end{document}